\documentclass{article}

\usepackage{arxiv}
\usepackage{array} 
\newcolumntype{L}{>{\let\newline\\\arraybackslash\hspace{0pt}}m{11cm}}
\usepackage{threeparttable} 
\usepackage{multirow}
\usepackage{epsfig}

\usepackage{amsmath}
\usepackage[utf8]{inputenc} 
\usepackage[T1]{fontenc}    
\usepackage{hyperref}       
\usepackage{url}            
\usepackage{booktabs}       
\usepackage{amsfonts}       
\usepackage{nicefrac}       
\usepackage{microtype}      
\usepackage{lipsum}
\usepackage{graphicx}
\graphicspath{ {./images/} }

\title{fkbma: An R Package for Detecting Tailoring Variables with Free-Knot B-Splines and Bayesian Model Averaging}

\author{
 Lara Maleyeff \\
 Department of Epidemiology, Biostatistics, and Occupational Health, McGill University, Montr\'eal, QC, CA \\
  \texttt{lara.maleyeff@mcgill.ca} \\
   \And
 Shirin Golchi \\
 Department of Epidemiology, Biostatistics, and Occupational Health, McGill University, Montr\'eal, QC, CA \\
  \And
 Erica E. M. Moodie \\
Department of Epidemiology, Biostatistics, and Occupational Health, McGill University, Montr\'eal, QC, CA \\
}

\begin{document}
\maketitle
\begin{abstract}
Precision medicine aims to optimize treatment by identifying patient subgroups most likely to benefit from specific interventions. To support this goal, we introduce fkbma, an R package that implements a Bayesian model averaging approach with free-knot B-splines for identifying tailoring variables. The package employs a reversible jump Markov chain Monte Carlo  algorithm to flexibly model treatment effect heterogeneity while accounting for uncertainty in both variable selection and non-linear relationships. fkbma provides a comprehensive framework for detecting predictive biomarkers, integrating Bayesian adaptive enrichment strategies, and enabling robust subgroup identification in clinical trials and observational studies. This paper details the statistical methodology underlying fkbma, outlines its computational implementation, and demonstrates its application through simulations and real-world examples. The package's flexibility makes it a valuable tool for precision medicine research, offering a principled approach to treatment personalization.
\end{abstract}


\noindent \textbf{Keywords:} Bayesian model averaging; Free-knot B-Splines, Precision medicine, Reversible jump MCMC, Predictive biomarkers

\section{Introduction}

Precision medicine aims to tailor treatments based on individual patient characteristics, with the dual goals of improving clinical outcomes and reducing exposure to ineffective therapies, side effects, or unnecessary costs. In clinical practice, there is increasing interest in the potential for improved treatment through biomarker-driven treatment strategies. These strategies allow clinicians to identify patients who are most likely to benefit from specific therapies, thereby personalizing care and optimizing resource allocation. Beyond clinical practice, precision medicine principles are being actively tested in clinical trials to enhance their efficiency. By focusing on biomarker-defined subgroups, these trials aim to reduce the number of participants exposed to ineffective treatments, thereby accelerating the development of targeted therapies.

A plethora of approaches to modeling treatment effect heterogeneity, whether in the context of randomized or non-experimental data, including but not limited to g-computation, g-estimation, A-learning, inverse probability of treatment weighting (with or without augmentation), outcome weighted learning, and more \cite{murphy2003optimal, robins2004optimal, zhao2012estimating, chakraborty2013statistical,tsiatis2019dynamic}. While many approaches explicitly rely on methodology that leverages modeling the treatment mechanism, others focus on flexibly modeling the outcome process so as to avoid model mis-specification and thus ensure consistency of estimated treatment effects. For example, machine learning techniques such as classification and regression trees (CART) and random forests, have been employed to identify treatment-effect heterogeneity in observational data \cite{breiman2001random, breiman2017classification,wager2018estimation}. These methods are particularly useful for discovering subgroups that may benefit differentially from one treatment as compared to an alternative, though they often lack the interpretability and statistical rigor of model-based approaches.

Bayesian methods have also been widely applied in for subgroup detection and treatment effect estimation. For instance, Bayesian additive regression trees (BART) have been used to model complex relationships between covariates and outcomes, providing a flexible framework for estimating heterogeneous treatment effects \cite{hill2020bayesian,chipman2010bart,hahn2020bayesian,logan2019decision}. Bayesian hierarchical models are particularly well-suited for precision medicine applications, as they allow for the incorporation of prior knowledge and provide a natural framework for uncertainty quantification.

Within the context of clinical trials, precision medicine is gaining traction through Bayesian adaptive enrichment designs which have ability to efficiently identify subgroups of patients with enhanced treatment effects, aligning with the goals of precision medicine. These designs leverage Bayesian methods to adaptively modify trial enrollment criteria based on accumulating data, focusing recruitment on subgroups that show the most promising treatment responses \cite{wang2007approaches,rosenblum2016group,mehta2014biomarker}. For instance, \cite{simon2013adaptive} proposed a Bayesian adaptive enrichment approach that uses posterior probabilities of treatment efficacy to refine subgroup definitions during the trial, ensuring resources are allocated to patients most likely to benefit. Basket trials, which evaluate a single treatment across multiple biomarker-defined subgroups, have also gained popularity for their ability to assess treatment effects in heterogeneous populations \cite{park2020overview}. Recent advancements, such as those by \cite{psioda2021bayesian}, have incorporated Bayesian model averaging (BMA) to address uncertainty in model selection, enabling robust decision-making through information borrowing across multiple models.

A key challenge in precision medicine is identifying subgroups of patients who are most likely to benefit from a particular treatment. While many biomarker-driven designs define patient subgroups \textit{a priori}, there is growing interest in more flexible approaches that incorporate biomarker identification during the analysis. For instance, \cite{park2022bayesian} proposed a Bayesian group sequential enrichment design that uses spike-and-slab priors for variable selection, allowing for the refinement of candidate biomarker space at each interim analysis. This method accommodates both categorical and continuous biomarkers, enabling the identification of treatment-sensitive subgroups. Similarly, \cite{liu2022bayesian} introduced a Bayesian adaptive design that models nonlinear relationships between biomarkers and treatment effects, offering a flexible framework for continuous biomarkers without fixed thresholds. However, this approach does not incorporate variable selection, limiting its utility in high-dimensional settings. \cite{maleyeff2024adaptive} addressed these limitations by developing a Bayesian adaptive enrichment design that identifies important predictive biomarkers from a larger set of candidates. Their approach uses free-knot B-splines to model the effects of continuous biomarkers and employs BMA to marginalize over all possible biomarker combinations. This flexible modeling framework allows for the identification of predictive variables and the assessment of treatment effects within biomarker-defined subgroups. Interim analyses enable early termination due to efficacy or futility, further enhancing trial efficiency.

Reversible jump Markov chain Monte Carlo (rjMCMC) methods have emerged as a powerful tool for BMA, enabling the exploration of complex model spaces with varying dimensions. Introduced by \cite{green1995reversible}, rjMCMC facilitates transitions between models of different sizes, such as those with varying numbers of covariates or parameters, by constructing a reversible Markov chain that maintains detailed balance across models. This approach is particularly useful in settings where model uncertainty is high, such as in variable selection or subgroup analysis, as it allows for the integration of posterior model probabilities into inference. Despite its flexibility, rjMCMC can be computationally intensive, and its performance often depends on the design of efficient proposal distributions to ensure adequate mixing across models \cite{brooks2003efficient}. 

Several R packages provide BMA functionality, though most are limited to models with simple linear or fixed terms. The BMA package \cite{raftery2005bma} is a foundational tool for BMA in R, designed primarily for linear regression. It identifies high-probability models by selectively sampling subsets of predictors and leverages Occam’s window for efficiency \cite{onorante2016dynamic}. Similarly, BAS \cite{clyde2018bas} extends BMA to generalized linear models, while BoomSpikeSlab \cite{scott2023package} specializes in models with sparse priors for high-dimensional predictors. However, these packages are unsuitable for models with non-linear effects, such as spline-based terms, where flexibility in selecting both the functional form and knot locations is essential. In contrast, packages like rstan and brms support a wider range of model structures, including non-linear and hierarchical models, and employ advanced MCMC algorithms such as Hamiltonian Monte Carlo (HMC) and the No-U-Turn Sampler (NUTS) \cite{burkner2017brms,rstan}. While these packages offer flexible modeling and MCMC sampling options, they do not perform BMA, limiting their utility for analysts seeking model averaging within complex structures.

We introduce \texttt{fkbma}, an R package that implements a rjMCMC sampler to identify important predictive biomarkers from a larger set of candidates. Building on the methods of \cite{maleyeff2024adaptive}, the package models the effects of continuous biomarkers using free-knot B-splines and employs BMA to marginalize over the space of all possible biomarker combinations. This approach reduces dependence on a single model specification and provides a robust framework for subgroup detection and treatment effect estimation. The rjMCMC algorithm used in \texttt{fkbma} is particularly well-suited for problems involving variable selection and flexible modeling. It allows for the exploration of a wide range of model structures, including those with non-linear effects, and provides a principled approach to model uncertainty. The package is designed to be applicable not only to precision medicine but also to any context where BMA and splines are relevant.

In Section \ref{sec:stat_med}, we provide an outline of the model specification, priors, and rjMCMC procedure, along with a discussion of the unique challenges of assessing convergence in this setting. Section \ref{sec:rjmcmc} describes the implementation of rjMCMC in the R package, and Section \ref{sec:example} provides examples of its usage. We conclude with a brief discussion in Section \ref{sec:disc}, highlighting the broader applicability of the methods and potential future extensions.

\section{Statistical Methods}
\label{sec:stat_med}
\subsection{Notation}
For $i=1,\dots,n$, let $Y_i$ be the outcome for individual $i$, $\mathbf{x}_i$ be a set of candidate continuous prognostic variables for $Y_i$, $\mathbf{z}_i$ be a set of candidate binary prognostic variables for $Y_i$, and $\mathbf{\tilde{x}}_i$,
$\mathbf{\tilde{z}}_i$ be sets of candidate continuous and binary predictive variables, respectively. Throughout, \textit{predictive} variables refer to those which interact with the exposure to modify exposure effects within specific subgroups, while a \textit{prognostic} variable directly influences the outcome regardless of interactions. Note that in the precision medicine literature, predictive variables may also be referred to as prescriptive or tailoring variables \cite{chakraborty2013statistical}. As our model was developed within the context of enrichment designs, we will conform with the terminology that is most common in this literature. For both binary and continuous variables, the set of candidate predictive variables is constrained to be a subset of the set of candidate prognostic variables. 

\subsection{Model Specification and Priors}
The \texttt{fkbma} R package implements a fully Bayesian model that flexibly accounts for continuous relationships using free-knot B-splines and performs BMA to incorporate model uncertainty. This approach allows for flexible modeling of non-linear associations while also identifying relevant predictors. The outcome \( Y_i \) is modeled using both continuous and binary covariates, where continuous variables are represented using splines to capture complex relationships.

For each individual \( i \), the expected outcome is modeled as:
\[
\mathbb{E}\left[Y_{i} \mid \mathbf{x}_i, \mathbf{z}_i, T_i\right] = \mu + \sum_{j=1}^{J} h_{1,j}(x_{ji}) + \boldsymbol{\beta}_1^\top \mathbf{z}_i + \left\{\phi + \sum_{j=1}^{\tilde{J}} h_{2,j}(\tilde{x}_{ji}) + \boldsymbol{\beta}_2^\top \mathbf{\tilde{z}}_i\right\} T_i + \epsilon_i,
\]
where \( \mu \) is the intercept, \( h_{1,j}(\cdot) \) and \( h_{2,j}(\cdot) \) are spline functions for continuous prognostic and predictive variables, respectively, and \( \boldsymbol{\beta}_1 \) and \( \boldsymbol{\beta}_2 \) represent the effects of binary prognostic and predictive variables. The term \( T_i \) is a binary indicator representing a factor of interest, which can take on different meanings depending on the context.

For example, in a randomized trial, \( T_i \) can represent treatment assignment, where \( T_i = 1 \) for treated individuals and \( T_i = 0 \) for controls. In this case, \( \phi \) captures the effect of treatment at reference levels of $\tilde{\mathbf{x}}$ and $\tilde{\mathbf{z}}$, and the terms \( h_{2,j}(\cdot) \) and \( \boldsymbol{\beta}_2 \) model treatment effect heterogeneity based on continuous and binary predictive variables. This formulation provides a flexible framework for modeling interactions and non-linear relationships while incorporating model uncertainty through BMA. By leveraging free-knot B-splines, the model can adaptively determine the appropriate complexity required to capture underlying patterns in the data.

In order to detect heterogeneous exposure effects, a common goal is to identify the effective subspace. Let $\boldsymbol{\gamma}(\mathbf{\tilde{x}},\mathbf{\tilde{z}}) = \phi + \sum_{j=1}^{\tilde{J}} h_{2,j}(\tilde{x}_{ji}) + \boldsymbol{\beta}_2^T \mathbf{\tilde{z}}$ be the variable-specific exposure effect or blip effect of the exposure \cite{robins1997causal}. The effective subspace can then be defined as:
\begin{equation}
\label{eq:eff_subspace}
    \mathcal{X}^* = \left\{\mathbf{\tilde{x}}, \mathbf{\tilde{z}}: P\left( \boldsymbol{\gamma}(\mathbf{\tilde{x}},\mathbf{\tilde{z}}) > 0 \mid \mathcal{D}\right)>1-\alpha\right\},
\end{equation}
where $\alpha$ is a pre-specified threshold. When used in the context of a Bayesian adaptive clinical trial, $\alpha$ may be tuned through simulation. For example, the effective subspace for patients who are responsive (i.e., $\boldsymbol{\gamma}(\mathbf{\tilde{x}},\mathbf{\tilde{z}}) > 0$ with high probability) to a novel treatment of hypertension may be found to be men over 60 years of age with a baseline systolic blood pressure greater than 140 mmHg and no history of diabetes.

The model incorporates a variety of priors to control the complexity of the included terms and the spline functions. The number of included terms (biomarker effects) follows a truncated Poisson distribution with parameter \( \lambda_1 \). Similarly, the number of knots in each spline term follows a truncated Poisson distribution with parameter \( \lambda_2 \). The model coefficients are assigned independent Normal priors with mean 0 and variance \( \sigma_B^2 \). The error term \( \epsilon_i \) is modeled as \( \epsilon_i \sim \mathcal{N}(0, \sigma_\epsilon^2) \), with an \(\text{Inverse-Gamma}(a_0, b_0)\) prior on \( \sigma_B \).

\subsection{Overview of the rjMCMC Procedure}

The core of the computation in \texttt{fkbma} relies on a custom reversible jump Markov Chain Monte Carlo (rjMCMC) procedure \cite{green1995reversible,maleyeff2024adaptive}. This allows for simultaneous exploration of models with varying dimensions, as both the number of terms (biomarker effects) and the complexity of continuous variables (number of spline knots) can change at each iteration.

\paragraph{Initialization} At the start of the rjMCMC procedure, the algorithm initializes several key components. Knot configurations for each continuous variable are set up, determining the initial placement of knots used for the spline terms. Simultaneously, the coefficient inclusion vector and the model coefficients are initialized. These coefficients include the intercept, exposure effects, binary variable effects, and spline coefficients, which form the baseline for subsequent updates in the MCMC iterations. Additionally, for models with Normally distributed outcomes, the individual-level heterogeneity parameter \( \sigma_\epsilon \) is initialized, with its prior following an Inverse-Gamma distribution. These initial values provide the starting point from which the rjMCMC sampler begins exploring the model space.

\paragraph{Iterative Steps} The procedure alternates between several key steps:

\begin{enumerate}
    \item \textbf{Spline Knot Updates}: For each spline term, the procedure can propose a change in knot location, addition of a knot (birth step), or removal of a knot (death step). This involves updating both the knot configuration and the associated spline coefficients. The birth or death of a knot uses rjMCMC moves, with the acceptance probability calculated based on the likelihood ratio, prior probabilities for the number and location of knots, and proposal distributions.
    \item \textbf{Biomarker Term Updates}: At each iteration, a biomarker term (either continuous or binary) may be added to or removed from the model, again using a birth or death step via rjMCMC. The acceptance probability for these moves is derived from the likelihood ratio, prior probabilities for model size, and the proposal distribution for the new term's coefficients. This step ensures that the model dynamically adjusts to include or exclude terms based on their relevance to the data.
    \item \textbf{Coefficient Updates}: The remaining model coefficients, including the intercept, exposure effects, and binary biomarker effects, are updated using Metropolis-Hastings (MH) steps. These updates propose small changes to the coefficients based on a random walk, with acceptance based on the likelihood and prior ratios.
    \item \textbf{Heterogeneity Parameter Update}: For Normally-distributed outcomes, the individual-level heterogeneity term \( \sigma_\epsilon \) is updated using a Gibbs sampling step, with a conjugate Inverse-Gamma prior.
\end{enumerate}

This iterative process is repeated for a large number of iterations, generating a sequence of states that characterizes the posterior distribution of all unknown parameters, including the set of included biomarkers, the number and location of spline knots, and the model coefficients. This allows \texttt{fkbma} to explore a large space of possible models while accounting for uncertainty in both the biomarker selection and the complexity of the spline terms.

The rjMCMC procedure ensures that the model space is efficiently explored, balancing model fit with simplicity, and dynamically adjusting both the number of included biomarkers and the spline terms to capture the most relevant exposure interactions. This methodology and computational framework allow \texttt{fkbma} to identify exposure-sensitive subgroups in a data-driven manner, making it a powerful tool for precision medicine trials. For more details, refer to \cite{maleyeff2024adaptive}.

\subsection{Convergence Diagnostics}

Assessing convergence in an rjMCMC sampler presents unique challenges, as individual parameters can appear and disappear across different model configurations throughout the sampling sequence. This variability complicates direct tracking of individual parameters and standard convergence diagnostics. To address this, we apply conventional MCMC diagnostics to parameters consistently included across all sub-models \cite{sisson2005transdimensional}. In our model, examples include the main exposure effect, $\phi$, which is present in all sub-models, as well as the tailored exposure effect for each individual. Although individual exposure effects may vary in dependence on different biomarkers throughout the rjMCMC sequence, the parameter itself remains well-defined, allowing for continuous trajectory tracking.

Convergence of the rjMCMC sampler is assessed using standard diagnostics, including trace plots for key model parameters. Other metrics, such as effective sample size and the Gelman-Rubin diagnostic ($\hat{R}$), are used to verify sufficient mixing of the chains, further confirming convergence \cite{gelman1992inference}.

\section{Reversible Jump MCMC Procedure: \texttt{{rj}MCMC}}
\label{sec:rjmcmc}

The rjMCMC algorithm is implemented in the function \texttt{rjMCMC}, which takes as input a data frame containing observations of a continuous outcome, binary exposure indicators, and candidate predictor variables. It implements the rjMCMC procedure described in \cite{maleyeff2024adaptive}, which generates posterior distributions by employing BMA and free-knot B-splines. The function is designed to handle complex models that include interactions between continuous and binary covariates, where the inclusion of model terms is subject to uncertainty, and the number and placement of knots in spline terms are unknown. The function allows the user to specify continuous and binary prognostic and predictive variables that should be included in the model. It employs a combination of MH updates for the model parameters and reversible jump steps for the model structure, including the addition and removal of knots in spline terms. Syntax for the function is:
\begin{verbatim}
rjMCMC(data, candsplinevars, candbinaryvars, candinter, outcome, 
        factor_var, mcmc_specs, prior_params)
\end{verbatim}
The mandatory arguments are \texttt{data}, a data frame containing columns for the outcome, exposure group, and candidate variables; \texttt{candsplinevars}, a character vector of names of continuous candidate prognostic variables that will be modeled with spline terms; \texttt{candbinaryvars}, a character vector of names of binary candidate prognostic variables; \texttt{candinter}, a character vector of names of candidate predictive variables (subset of \texttt{candsplinevars} and \texttt{candbinaryvars}); \texttt{outcome}, a string containing the column name of the outcome variable; and \texttt{factor\_var}, a string containing the column name of the exposure variable (e.g., treatment). Optional parameters are \texttt{mcmc\_specs}, specifications for the MCMC procedure; and \texttt{prior\_params}, prior parameters for model terms. Table \ref{tab:input_params} describes the parameters \texttt{mcmc\_specs} and  \texttt{prior\_params} and their default values in more detail.

\begin{table}
\caption{Input parameters for \texttt{fkbma} package.}
\label{tab:input_params}
\centering
\begin{tabular}{lLc}
\textbf{Parameter Name} & \textbf{Description} & \textbf{Default} \\ \hline
\texttt{mcmc\_specs} & (MCMC Specifications) A list with elements: \\ 

    \ \ \ \ \texttt{iter} & Number of total iterations per chain & 4000 \\
    \ \ \ \ \texttt{warmup} & Number of burnin iterations & \texttt{iter/2} \\
    \ \ \ \ \texttt{thin} & Thinning parameter & 1 \\
    \ \ \ \ \texttt{chains} & Number of MCMC chains & 4 \\
    \ \ \ \ \texttt{sigma\_v} & Proposal variance for ``jump" terms & 0.1 \\
    \ \ \ \ \texttt{bma} & Boolean for Bayesian model averaging & \texttt{TRUE} \\ \hline

\texttt{prior\_params} & (Prior Distribution Parameters) A list with elements: \\

    \ \ \ \ \texttt{lambda\_1} & Prior for number of terms in the model & 0.1 \\
    \ \ \ \ \texttt{lambda\_2} & Prior for number of knots in each spline & 1 \\
    \ \ \ \ \texttt{a\_0} & Shape parameter for inverse gamma prior on variance & 0.01 \\
    \ \ \ \ \texttt{b\_0} & Rate parameter for inverse gamma prior on variance & 0.01 \\
    \ \ \ \ \texttt{degree} & Degree of B-splines & 3 \\
    \ \ \ \ \texttt{k\_max} & Maximum number of knots for each spline term & 9 \\
    \ \ \ \ \texttt{w} & Window for proposing knot location changes & 1 \\
    \ \ \ \ \texttt{sigma\_B} & Prior standard deviation for model coefficients & \( \sqrt{20} \) \\ \hline
\end{tabular}
\end{table}

The output of the function is an object of class \texttt{rjMCMC} that includes the posterior samples of the exposure effect, binary parameters, spline coefficients, model variance, and the number of knots in each spline term. It also includes the acceptance rates for proposed moves.

\begin{table}
\centering
\caption{Summary of plot function options based on \texttt{sample\_type}, \texttt{plot\_type}, and \texttt{effect\_type}.}
\begin{threeparttable}
\begin{tabular}{|l|l|l|p{7.5cm}|}
\hline
\textbf{\texttt{sample\_type}} & \textbf{\texttt{plot\_type}} & \textbf{\texttt{effect\_type}} & \textbf{Description} \\ \hline
\multirow{2}{*}{\texttt{estimand}\tnote{1}} & \texttt{hist/trace} & \texttt{outcome} & For \texttt{hist}, plots the posterior distribution (e.g., coefficient of \texttt{Z\_1}) without exposure interaction; \texttt{trace} plots the MCMC sampling trajectory of this main effect. \\ \cline{2-4} 
 & \texttt{hist/trace} & \texttt{exposure\_effect} & For \texttt{hist}, plots the posterior distribution of the interaction effect (e.g., coefficient of \texttt{Z\_1:trt}); \texttt{trace} plots the MCMC trajectory of the exposure interaction effect. \\ \hline
\multirow{4}{*}{\texttt{fitted/predictive}} & \texttt{cred} & \texttt{outcome} & Plots the mean and credible interval for \texttt{fitted} or \texttt{predictive} outcome values on the y-axis against biomarker values on the x-axis. \\ \cline{2-4} 
 & \texttt{cred} & \texttt{exposure\_effect} & Plots the \texttt{fitted} or \texttt{predictive} exposure effect, showing the $\gamma(\tilde{\mathbf{x}},\tilde{\mathbf{y}})$ function or the effect of exposure across covariate patterns with credible intervals. \\ \cline{2-4} 
 & \texttt{hist/trace} & \texttt{outcome} & Plots the trajectory of one person's \texttt{fitted} or \texttt{predictive} outcome value over time or across MCMC iterations. \\ \cline{2-4} 
 & \texttt{hist/trace} & \texttt{exposure\_effect} & For \texttt{fitted} or \texttt{predictive} values, plots the trajectory of one person’s exposure effect over time or across MCMC iterations. \\ \hline
\end{tabular}
\begin{tablenotes}
\item[1] For parameters \texttt{intercept}, \texttt{factor\_var}, and \texttt{sigma}, \texttt{effect\_type} does not apply as they have no interaction effect.
\end{tablenotes}
\end{threeparttable}
\label{tab:plotoptions}
\end{table}

The \texttt{rjMCMC} object is supported by the standard functions: \texttt{summary}, \texttt{print}, \texttt{predict},  \texttt{fitted}, \texttt{coef}, and  \texttt{plot}. The \texttt{summary} function provides a detailed summary of model information, including the fitted formula, convergence diagnostics, MCMC sampler arguments, and posterior estimates. It displays posterior means, credible intervals, and posterior inclusion probabilities for binary parameters and spline terms, as well as residual variance estimates. The \texttt{print} function serves as a wrapper for \texttt{summary}, providing a simplified method for printing summary results. The \texttt{predict} function generates posterior predictive values for new observations by drawing new outcomes from the model, incorporating fixed effects, spline terms, and binary parameters while also accounting for residual variance. In contrast, the \texttt{fitted} function  generates posterior fitted values for the observed data, which represent the model’s expected outcomes based on the estimated parameters but without additional uncertainty from residual variation. The \texttt{plot} function provides flexible visualization tools for assessing the posterior distributions, convergence, and exposure effects generated by the \texttt{rjMCMC} model. Users can specify different types of plots, such as trace plots, histograms, and credible intervals, designed to show the main effect of exposure ($\phi$), individual exposure effects, or predictive distributions for continuous and binary covariates. The function output options, summarized in Table~\ref{tab:plotoptions}, allow for customized visualization based on desired sample type (\texttt{estimand}, \texttt{fitted}, or \texttt{predictive}), plot type (\texttt{hist}, \texttt{trace}, or \texttt{cred}), and effect type (\texttt{outcome} or \texttt{exposure\_effect}). The \texttt{sample\_type} argument specifies whether to visualize posterior distributions of model parameters (\texttt{estimand}), fitted values for the observed data for a specific individual or covariate pattern (\texttt{fitted}), or out-of-sample predictions for a specific individual or covariate pattern (\texttt{predictive}). The \texttt{plot\_type} argument controls the visualization format, offering histograms for posterior distributions (\texttt{hist}), trace plots for MCMC trajectories (\texttt{trace}), or credible interval plots for fitted or predictive values (\texttt{cred}). The \texttt{effect\_type} argument determines the focus of the visualization, with options to examine main effects of covariates or biomarkers on the outcome (\texttt{outcome}) or interaction effects, such as the effect of exposure (e.g., treatment) across different levels of covariates or biomarkers (\texttt{exposure\_effect}). For example, users can plot the posterior distribution of the coefficient for \texttt{Z\_1} (\texttt{outcome}) or the interaction effect between \texttt{Z\_1} and treatment (\texttt{exposure\_effect}), providing insights into both direct and moderated effects. This flexibility facilitates thorough exploration of model results and visual assessment of parameter behavior across subgroups defined by biomarkers. The \texttt{coef} function extracts posterior mean coefficients associated with the intercept, exposure main effect, and binary parameters. 

The \texttt{rjMCMC} object is also supported by additional non-standard functions. To assess the model's variable selection, the \texttt{pip} function computes and returns posterior inclusion probabilities for all candidate variables. For assessing parameter uncertainty, the \texttt{credint} function calculates and displays credible intervals for model parameters. The functions \texttt{fittedTrtEff} and \texttt{predictMainEffect} compute fitted and predictive MCMC samples for the effect of exposure for each individual's biomarker values. The function \texttt{getEffectiveSubspace} identifies the effective subspace, or the individuals who are expected to benefit from exposure. Finally, the \texttt{rhats} function computes the Gelman-Rubin convergence diagnostic \cite{gelman1992inference} ($\widehat{R}$) for individual exposure effects, intercept, and exposure effect parameters, providing key information on MCMC convergence.

\section{Example Usage}
\label{sec:example}

Below is an example usage of the \texttt{rjMCMC} function to fit a model with one continuous spline variable (\texttt{X\_1}) and five binary variables (\texttt{Z\_1}, \texttt{Z\_2}, \texttt{Z\_3}, \texttt{Z\_4}, \texttt{Z\_5}). 
In this example, data are generated for a sample of size 1000. The continuous variable \texttt{X\_1} is drawn from a uniform distribution between 0 and 1. Five binary variables (\texttt{Z\_1} through \texttt{Z\_5}) are generated from independent Bernoulli distributions with different probabilities of success: \texttt{Z\_1} and \texttt{Z\_5} with a probability of 0.35, \texttt{Z\_2} with a probability of 0.5, \texttt{Z\_3} with a probability of 0.65, and \texttt{Z\_4} with a probability of 0.2. The treatment variable \texttt{trt} is also binary and follows a Bernoulli distribution with a success probability of 0.5. The response variable \texttt{Y} is generated as a  function of \texttt{X\_1} and \texttt{Z\_1}, and their interactions with \texttt{trt}, with an added noise term drawn from a normal distribution with mean 0 and standard deviation 0.5:
\[
Y = 2 \cdot Z_1 + 2 \cdot X_1 + 2 \cdot Z_1 \cdot \text{trt} + \cos(2 \pi X_1) \cdot \text{trt} + \epsilon,
\]
where \(\epsilon \sim \mathcal{N}(0, 0.5^2)\).

\begin{verbatim}
n <- 1000
data <- data.frame(
  X_1 = runif(n, 0, 1),
  Z_1 = rbinom(n, 1, 0.35),
  Z_2 = rbinom(n, 1, 0.5),
  Z_3 = rbinom(n, 1, 0.65),
  Z_4 = rbinom(n, 1, 0.2),
  Z_5 = rbinom(n, 1, 0.35),
  trt = rbinom(n, 1, 0.5)
)
data$Y = 2 * data$Z_1 + 2 * data$X_1 + 2 * data$Z_1 * data$trt + 
    cos(data$X_1*2*pi) * data$trt + rnorm(n, 0, 0.5)
\end{verbatim}

The generated dataset is then used to fit a model using the \texttt{rjMCMC} function, incorporating a continuous spline for \texttt{X\_1}, the binary variables, and the interaction of the biomarkers with treatment. The default settings for \texttt{mcmc\_specs} and \texttt{prior\_params} are used.

\begin{verbatim}
candsplinevars <- c("X_1")
candbinaryvars <- paste0("Z_", 1:5)
candinter <- c(candsplinevars, candbinaryvars)

results <- rjMCMC(data, candsplinevars, candbinaryvars, candinter, outcome = "Y",
           factor_var = "trt")
\end{verbatim}

The total runtime was 5 minutes 50 seconds on a MacBook Pro (13-inch, 2020, Intel, macOS Ventura 13.4.1). The results include posterior distributions for the treatment effect, binary and spline variables, and diagnostic summaries.

\begin{verbatim}
summary(results)

# Model Information:
# Formula:  
# Y ~ fbs(X_1) + Z_1 + Z_2 + Z_3 + Z_4 + Z_5 + trt + 
#     fbs(X_1):trt + Z_1:trt + Z_2:trt + Z_3:trt + Z_4:trt + Z_5:trt 
# Note: fbs() indicates a free-knot B-spline.
# Data: results$data_fit
# Number of observations: 1000

# MCMC Sampler Arguments:
#   - iter: 4000
#   - warmup: 2000
#   - thin: 1
#   - chains: 4

# Parameter Estimates:

# Non-spline Parameters:
#           Estimate Est.Error l-95% CI u-95% CI Eff.Sample  Rhat   PIP
# intercept    0.009     0.087   -0.173    0.173    323.233 1.003 1.000
# trt          1.227     0.152    0.943    1.532    288.461 1.006 1.000
# Z_1          2.042     0.048    1.948    2.136   5367.806 1.000 1.000
# Z_1:trt      1.895     0.068    1.760    2.029   6220.389 1.000 1.000
# Z_5          0.000     0.002    0.000    0.000         NA    NA 0.003
# Z_3          0.000     0.002    0.000    0.000         NA    NA 0.002
# Z_4          0.000     0.000    0.000    0.000         NA    NA 0.000
# Z_2          0.000     0.000    0.000    0.000         NA    NA 0.000
# Z_2:trt      0.000     0.000    0.000    0.000         NA    NA 0.000
# Z_3:trt      0.000     0.000    0.000    0.000         NA    NA 0.000
# Z_4:trt      0.000     0.000    0.000    0.000         NA    NA 0.000
# Z_5:trt      0.000     0.000    0.000    0.000         NA    NA 0.000

# PIP = posterior inclusion probability

# Gaussian Family Specific Parameters:
#       Estimate Est.Error l-95% CI u-95% CI Eff.Sample Rhat PIP
# sigma    0.518     0.012    0.496    0.542   30463.09    1   1

# Posterior Inclusion Probabilities for Splines:
#     X_1 X_1:trt 
#       1       1
\end{verbatim}

Based on the analysis using the \texttt{rjMCMC} function, the model correctly identified \texttt{Z\_1} and \texttt{X\_1} as predictive variables, as indicated by their posterior inclusion probabilities (PIPs) being equal to 1. This result confirms the substantial contribution of these variables to the model's predictive capacity. For spline variables like \texttt{X\_1}, where parameter definitions vary across MCMC iterations, the exact effect is challenging to report in a numerical summary alone. Instead, visual examination is recommended to better understand the nature and structure of the estimated effects. Looking at the MCMC diagnostics, the effective sample sizes (\texttt{Eff.Sample}) are generally high for included parameters, such as \texttt{Z\_1:trt} and \texttt{Z\_1}, suggesting sufficient mixing and precision in parameter estimates. The \texttt{Rhat} values for these parameters are close to 1, indicating good convergence. However, for some parameters like the treatment variable (\texttt{trt}), lower effective sample sizes (e.g., \texttt{Eff.Sample} of 288.461) suggest that this parameter may benefit from additional MCMC samples for improved stability.

\begin{verbatim}
plot(results)
plot(results, sample_type = "predictive")
# Automatically setting variables to be continuous model variables 
# with pip > pip_cutoff.
# Automatically setting facet_by to be binary model variables with pip > pip_cutoff.
# The following variables were not provided and will be set to 0: Z_2, Z_3, Z_4, Z_5
\end{verbatim}

\begin{figure}
    \centering
    \begin{minipage}{0.49\textwidth}
        \centering
        \includegraphics[width=\textwidth]{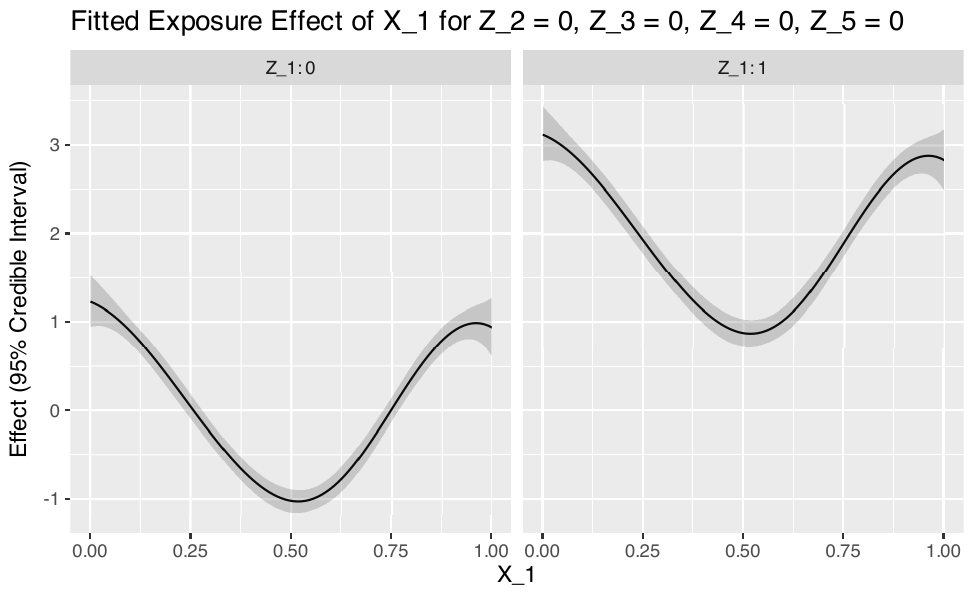} 
        \caption*{(A) Fitted treatment effect across biomarker \texttt{X\_1}, stratified by \texttt{Z\_1}.}
    \end{minipage}
    \hfill
    \begin{minipage}{0.49\textwidth}
        \centering
        \includegraphics[width=\textwidth]{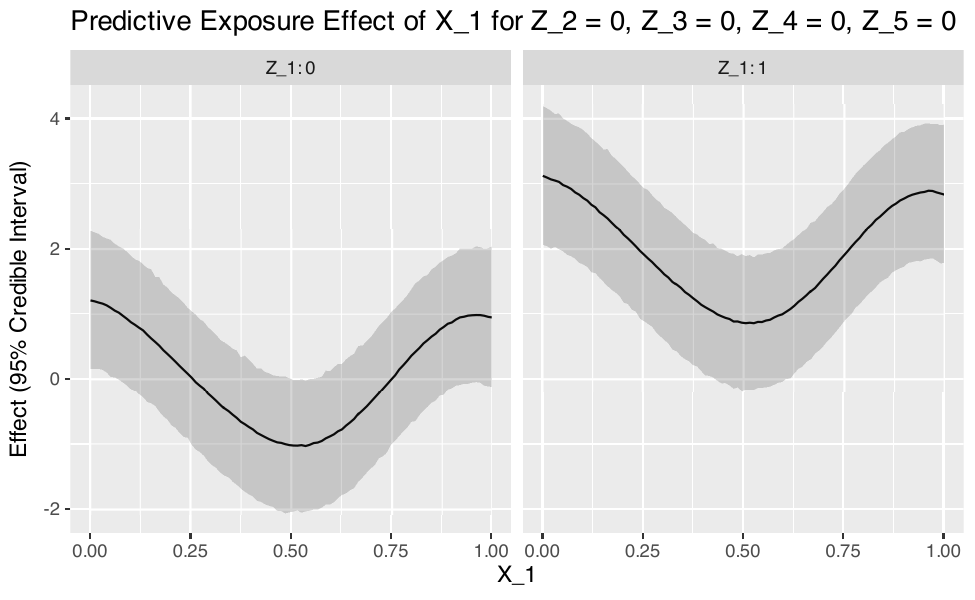} 
        \caption*{(B) Predictive treatment effect across biomarker \texttt{X\_1}, stratified by \texttt{Z\_1}.}
    \end{minipage}
    \caption[Fitted and predictive treatment effects based on data example.]{Fitted (A) and predictive (B) treatment effects based on continuous biomarker \texttt{X\_1}, stratified by binary covariate \texttt{Z\_1}. Panel A shows the estimated fitted values for the effect of treatment, while panel B shows the predictive values, with treatment effects shown across the range of \texttt{X\_1} for each level of \texttt{Z\_1}.}
    \label{fig:treatment_effects}
\end{figure}

By default, the \texttt{plot} function visualizes point-wise credible intervals for the fitted or predictive values of the treatment effect, displaying results for all predictive continuous variables stratified by predictive binary variables. Fitted values are computed as the posterior mean of the model's predictions for the observed data, conditional on the estimated parameters and covariates. These values reflect the model's fit to the data and are typically used to assess how well the model captures the observed relationships. In contrast, predictive values are computed as the posterior predictive distribution, which incorporates additional uncertainty by generating predictions for new data points, marginalizing over the posterior distribution of the parameters. This added uncertainty arises because predictive values account for both parameter uncertainty and inherent variability in the data-generating process. In this case (Figure \ref{fig:treatment_effects}), the plot represents \texttt{X\_1} versus fitted (A) or predictive (B) values, stratified by \texttt{Z\_1}. As anticipated, the credible intervals for the predictive values are wider than those for the fitted values, reflecting the increased uncertainty in out-of-sample predictions compared to in-sample fitted values. This distinction highlights the importance of considering both fitted and predictive values when evaluating model performance and generalizability.

The relationship between \texttt{X\_1} and the treatment effect exhibits an upward parabolic pattern: treatment appears less effective or even harmful for moderate values of \texttt{X\_1} and shows greater effectiveness at both high and low extremes of \texttt{X\_1}. Using a threshold of $\alpha=0.025$, a decision rule emerges: treat all individuals with \texttt{Z\_1} = 1, while for those with \texttt{Z\_1} = 0, apply treatment only if \texttt{X\_1} is outside the range (0.23, 0.77). This visually driven treatment rule highlights how stratification by both continuous and binary predictors can guide effective, targeted intervention strategies.

\begin{verbatim}
effective_subsapce = getEffectiveSubspace(results, alpha = 0.025)
# Effective subspace descriptions:

# Z_1 = 0: X_1 in [0, 0.23] or [0.77, 1], Z_1 = 1: X_1 in [0, 1]
\end{verbatim}

The function \texttt{getEffectiveSubspace} prints a summary of the effective subspace  using a significance cutoff of $\alpha$. The function returns the $\alpha$-quantile of the posterior distribution of the treatment effect for each individual, as well as an indicator variable for whether each $\alpha$-quantile is greater than $0$. For $\alpha=0.025$, this function confirms that the effective subspace consists of individuals where either \texttt{Z\_1} = 1, or \texttt{Z\_1} = 0 with \texttt{X\_1}  falling outside the range (0.23, 0.77).

\begin{verbatim}
plot(results, plot_type="trace")
# Automatically setting variables to be continuous model variables 
# with pip > pip_cutoff.
# Automatically setting facet_by to be binary model variables with pip > pip_cutoff.
# The following variables were not provided and will be set to 0: Z_2, Z_3, Z_4, Z_5
\end{verbatim}

\begin{figure}
    \centering
    \includegraphics[width=0.8\textwidth]{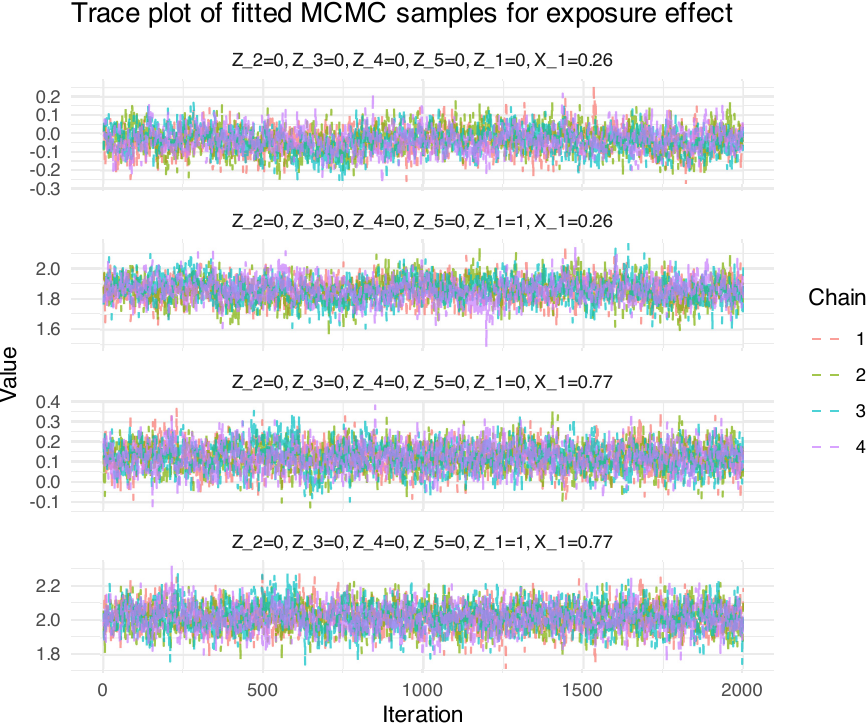}
    \caption[Trace plots showing the MCMC sampling trajectories for the data example.]{Trace plot showing the MCMC sampling trajectories for the treatment effect, stratified by values of the continuous biomarker \texttt{X\_1} and binary covariate \texttt{Z\_1}. Here, the \texttt{plot} function automatically selects the 0.25 and 0.75-quantiles of the continuous biomarker \texttt{X\_1} to represent low and high values, respectively. In this case, these quantiles happen to align closely with the predefined cutoffs for the effective subspace. For binary biomarkers, the function automatically identifies the two possible values (0 and 1) and generates trajectory plots for both cases. This visualization allows for a clear assessment of the treatment effect's stability and convergence across varying levels of \texttt{X\_1} and \texttt{Z\_1}.}
    \label{fig:trace_plot}
\end{figure}

The trace plot (Figure~\ref{fig:trace_plot}), generated using the option \texttt{plot\_type="trace"}, displays the MCMC sampling trajectories for the treatment effect, stratified by values of the continuous biomarker \texttt{X\_1} and the binary covariate \texttt{Z\_1}. In this case, the function will plot the trajectories for individuals with varying biomarker values (biomarkers automatically selected based on their PIPs). For continuous-valued biomarkers, the function selects the 0.25 and 0.75 quantiles to represent low and high values, respectively. In this case, these quantiles happen to align closely with the predefined cutoffs for the effective subspace. For binary biomarkers, the function automatically identifies the two possible values (0 and 1) and generates trajectory plots for both cases. This visualization provides a clear assessment of the treatment effect's stability and convergence across different levels of the continuous biomarker \texttt{X\_1} and the binary covariate \texttt{Z\_1}, offering insights into the sampler performance. The chains demonstrate good mixing, as evidenced by their smooth trajectories across iterations without observable trends, indicating that the MCMC sampler effectively explores the posterior distribution for the treatment effect within each subgroup.

\section{Conclusion}
\label{sec:disc}

The \texttt{fkbma} R package provides a powerful and flexible framework for identifying predictive biomarkers and estimating treatment effects in precision medicine and beyond. By combining  BMA and free-knot B-splines in a custom rjMCMC sampler, the package addresses the challenges of variable selection, non-linear modeling, and model uncertainty in a principled manner. This approach not only reduces dependence on a single model specification but also enables robust subgroup detection and exposure effect estimation, making it particularly well-suited for both clinical trials and observational studies. However, the utility of \texttt{fkbma} extends far beyond this domain. Its integration of BMA and flexible spline-based modeling makes it applicable to any field where similar challenges of variable selection and non-linear effects arise.

The rjMCMC algorithm implemented in \texttt{fkbma} is a key strength, allowing users to explore a broad space of model structures and incorporate uncertainty into their inferences. This flexibility is critical for applications where the underlying data-generating process is complex or poorly understood. Furthermore, the package’s ability to model continuous biomarkers using free-knot B-splines ensures that non-linear relationships are captured without over-reliance on restrictive parametric assumptions. This feature is particularly valuable in settings where biomarkers or covariates exhibit threshold effects or other non-linear patterns, which are common in many scientific and industrial contexts.

Looking ahead, we plan to extend the capabilities of \texttt{fkbma} beyond interactions with an exposure, allowing it to fit models of more general forms. This will enhance its applicability across a broader range of statistical and clinical settings. Additionally, we will add trial functionality for Bayesian adaptive designs to allow researchers to simulate trial designs, tune parameters to achieve desired operating characteristics, and easily assess decision rules. To this end, we aim to support interim analyses in adaptive clinical trial designs by incorporating early stopping criteria based on efficacy or futility within specific subgroups. This enhancement will improve trial efficiency and reduce patient exposure to ineffective treatments. Furthermore, we plan to implement more extensive design simulation tools, enabling users to optimize trial parameters and evaluate model performance across diverse clinical scenarios. These developments will further strengthen \texttt{fkbma}’s utility for precision medicine and its broader applicability to other fields.

In summary, \texttt{fkbma} represents an advancement in the analysis of predictive biomarkers and treatment effects, offering a robust and flexible framework for addressing complex modeling challenges. Its integration of rjMCMC, BMA, and free-knot B-splines makes it a valuable tool not only for precision medicine but also for any field requiring variable selection, non-linear modeling, and uncertainty quantification. As we continue to expand its capabilities, we anticipate that \texttt{fkbma} will become an increasingly versatile resource for researchers and practitioners across a wide range of disciplines.

\newpage

\section*{Acknowledgements}
\noindent Lara Maleyeff acknowledges the support of the Canadian Network for Statistical Training in Trials (CANSTAT) Postdoctoral Fellowship during the preparation of this research. Shirin Golchi is a Fonds de Recherche du Québec, Santé, Chercheuse-boursière (Junior 1) and acknowledges support from a Natural Sciences and Engineering Council of Canada (NSERC) Discovery Grant, Canadian Statistical Sciences Institute, and the Fonds de Recherche du Québec, Nature et technologies (FRQNT-NSERC NOVA). Erica E. M. Moodie is a Canada Research Chair (Tier 1) in Statistical Methods for Precision Medicine and acknowledges the support of a Chercheur de Mérite Career Award from the Fonds de Recherche du Québec, Santé. This work was supported by the National Institute of Mental Health of the National Institutes of Health under Award Number R01 MH114873. The content is solely the responsibility of the authors and does not necessarily represent the official views of the National Institutes of Health.

\section*{Disclosure statement}
\noindent No potential conflict of interest was reported by the authors.

\section*{Data availability statement}
\noindent Simulated data used in Section \ref{sec:example} are available in the R package \texttt{fkbma}.

\bibliographystyle{unsrt}  
\bibliography{main}  

\end{document}